\documentclass[aps,prx,reprint,showpacs,superscriptaddress]{revtex4-2}

\usepackage{graphicx}
\usepackage{dcolumn}
\usepackage{bm}
\usepackage{amsmath,amssymb,amsfonts}
\usepackage{clrscode}

\usepackage{multirow}
\usepackage{makecell}
\usepackage{hyperref}

\usepackage{textcomp}
\usepackage{epstopdf}
\usepackage{bm}
\usepackage{float}
\usepackage{threeparttable}
\usepackage{caption}
\usepackage{array}
\usepackage{setspace}
\usepackage{booktabs} 
\usepackage{gensymb}
\usepackage{multirow}
\makeatletter
\let\newfloat\newfloat@ltx
\makeatother
\usepackage{algorithm}
\usepackage{algorithmic}
\usepackage{color}
\usepackage{mathrsfs}
\usepackage{subfig}
\usepackage{siunitx}
\usepackage{lineno} 
\usepackage[draft]{changes}

\begin{document}

\preprint{APS/123-QED}

\title{Data-driven modeling of Landau damping by physics-informed neural networks}

\author{Yilan Qin}
\affiliation{School of Information Science and Engineering, Fudan University, Shanghai, 200433, China}
\affiliation{Key Laboratory for Information Science of Electromagnetic Waves (MoE), Fudan University, Shanghai, 200433, China}
\author{Jiayu Ma}
\affiliation{School of Information Science and Engineering, Fudan University, Shanghai, 200433, China}
\affiliation{Key Laboratory for Information Science of Electromagnetic Waves (MoE), Fudan University, Shanghai, 200433, China}
\author{Mingle Jiang}
\affiliation{School of Information Science and Engineering, Fudan University, Shanghai, 200433, China}
\affiliation{Key Laboratory for Information Science of Electromagnetic Waves (MoE), Fudan University, Shanghai, 200433, China}
\author{Chuanfei Dong}
\email{dcfy@bu.edu}
\affiliation{Department of Astronomy, Boston University, Boston, Massachusetts 02215, USA}
\affiliation{Princeton Plasma Physics Laboratory and Department of Astrophysical Sciences, Princeton University, Princeton, New Jersey 08540, USA}
\author{Haiyang Fu}
\email{haiyang\_fu@fudan.edu.cn}
\affiliation{School of Information Science and Engineering, Fudan University, Shanghai, 200433, China}
\affiliation{Key Laboratory for Information Science of Electromagnetic Waves (MoE), Fudan University, Shanghai, 200433,  China}
\author{Liang Wang}
\email{liang.wang@princeton.edu}
\affiliation{Princeton Plasma Physics Laboratory and Department of Astrophysical Sciences, Princeton University, Princeton, New Jersey 08540, USA}
\author{Wenjie Cheng}
\affiliation{School of Information Science and Engineering, Fudan University, Shanghai, 200433, China}
\affiliation{Key Laboratory for Information Science of Electromagnetic Waves (MoE), Fudan University, Shanghai, 200433, China}
\author{Yaqiu Jin}
\affiliation{School of Information Science and Engineering, Fudan University, Shanghai, 200433, China}
\affiliation{Key Laboratory for Information Science of Electromagnetic Waves (MoE), Fudan University, Shanghai, 200433, China}

\date{\today}

\begin{abstract}
Kinetic approaches are generally accurate in dealing with microscale plasma physics problems but are computationally expensive for large-scale or multiscale systems. One of the long-standing problems in plasma physics is the integration of kinetic physics into fluid models, which is often achieved through sophisticated analytical closure terms. In this paper, we successfully construct a multi-moment fluid model with an implicit fluid closure included in the neural network using machine learning. The multi-moment fluid model is trained with a small fraction of sparsely sampled data from kinetic simulations of Landau damping, using the physics-informed neural network (PINN) and the gradient-enhanced physics-informed neural network (gPINN). The multi-moment fluid model constructed using either PINN or gPINN reproduces the time evolution of the electric field energy, including its damping rate, and the plasma dynamics from the kinetic simulations. In addition, we introduce a variant of the gPINN architecture, namely, gPINN$p$ to capture the Landau damping process. Instead of including the gradients of all the equation residuals, gPINN$p$ only adds the gradient of the pressure equation residual as one additional constraint. Among the three approaches, the gPINN$p$-constructed multi-moment fluid model offers the most accurate results. This work sheds light on the accurate and efficient modeling of large-scale systems, which can be extended to complex multiscale laboratory, space, and astrophysical plasma physics problems.
\end{abstract}


\maketitle
\section{Introduction}

Microscale kinetic physics is crucial for accurately modeling many laboratory, space, and astrophysical systems~\cite{Daughton2011,Raymond2018,Dong2018,WangL2018,Dong2019,Fiuza2020,Li2023,Yang2023}. Unfortunately, for large-scale systems, the first-principle method, which is based on the direct numerical treatment of the kinetic equations, frequently incurs computational costs that are unaffordably expensive. To mitigate the computational cost of kinetic models, numerous attempts have been made to incorporate kinetic physics into the fluid framework that evolves a finite number of fluid moment equations constructed by taking velocity moments of the kinetic Vlasov equation~\cite{Hakim2008,Huang2019,Wang2020}. In the area of plasma physics, one profound attempt is the Landau-fluid models pioneered by Hammett and Perkins~\cite{1990Fluid}, who derived analytical closure relations for the truncated plasma fluid equations by matching the exact linear response associated with Landau damping in a collisionless electrostatic plasma. A lengthy series of works have gone into devising variants of the fluid closures in different regimes that greatly determine the validity and accuracy of the resulting models~\cite{Hunana2019}. Unfortunately, one major difficulty in constructing the fluid closures in collisionless plasmas is that it typically requires very nontrivial physical and mathematical analyses applied to the specific regime. With the quick development of artificial intelligence in the past decade, naturally, the question arises: Can machine learning assist in completing this challenging task by exploring the kinetic simulation data?

Indeed, using conventional artificial neural networks (ANNs) for the discovery of fluid closures in collisionless plasmas has been an active area of research. The earliest attempt was perhaps made by~\citet{2020Machine}, who trained a multilayer perceptron (MLP), a convolutional neural network (CNN), a discrete Fourier transform (DFT) network to learn the Hammett-Perkins closure. However, to the authors' knowledge, this and the subsequent attempts relied on training data from Landau fluid simulations with a known closure relation. Promising progress has been reported by~\citet{laperre2022} who used an MLP and a gradient boosting regressor to synthesize a local mapping from local information to local plasma pressure tensor and heat flux, using kinetic simulation data of a two-dimensional (2D) magnetic reconnection problem as the input. In their work, nonlocal closures were not investigated, and the mapping differs from conventional closure concepts where the plasma pressure is used as an input.

One common issue in applying the traditional ANN to the discovery of physical relations is its strong reliance on large datasets and slow convergence since the complex underlying physical constraints are not properly imposed. As a remedy, machine learning techniques and methods such as symbolic regression~\cite{bongard2007,schmidt2009distilling}, sparse regression~\cite{Brunton2016,rudy2017data,long_pde-net_2019}, and the physics-informed neural network (PINN)~\cite{2019Physics,2020Hidden} have been developed. In terms of theoretical plasma physics, some attempts have been recently made by distilling the data and selecting appropriate physical terms from a library of candidate terms. For instance,~\citet{Alves2022} explored sparsity-based model-discovery techniques in Ref.~\cite{rudy2017data} to discover multi-fluid and magnetohydrodynamic equations from the kinetic simulation data. Modified PDE-Net (mPDE-Net, where PDE refers to partial differential equation) has also been used to discover multi-moment fluid equations together with an explicit heat flux closure from kinetic simulation data~\cite{Xiong2019,Cheng2022}. However, such library-based frameworks rely on pre-defined candidate terms that are not always known or well understood. Among these venues, PINN is possibly one of the most influential examples. The physical partial differential equation (PDE) residuals are incorporated into the loss function of the neural network as regularization, transforming the process of solving PDEs into an optimization problem by constraining the space of permissible solutions. Since its introduction by~\citet{2019Physics}, PINN and its variants have been widely applied to fluid dynamics, plasma physics, electromagnetics, and many more areas~\cite{mathews2020uncovering,wang2021data,lou2021physics,cai2021flow,YYzhang2022}.
One remarkable improvement of PINN was made by~\citet{yu2022gradient}, who added additional gradient loss terms to construct the gradient-enhanced physics-informed neural network (gPINN) to improve the accuracy for large-gradient shock-wave physics. 

This work aims to explore the feasibility and effectiveness of capturing the hidden fluid closure using PINN without prescribing the form of the closure itself. A library of explicit candidate terms in the closure relations would not be necessary. The key point here is to use the kinetic simulation data that contains the complete closure information as the training datasets, and use fluid moment equations to constrain the training process. The trained neural network then embeds the closure
information implicitly and can be used to close the multi-moment fluid equation system and incorporate desirable kinetic physics. As a first but critical step, we will use the example of Landau damping in a collisionless, electrostatic plasma, which is one of the most fundamental kinetic processes in a variety of plasmas. We will explore the performance of the original PINN and its variants, in particular, the gradient-enhanced PINN (gPINN), in capturing the hidden fluid closure that can reproduce the Landau damping process. The results of this study could be extended to other more complex problems and be combined with more sophisticated, more general approaches.

\begin{figure*}[htbp]
\centering
\includegraphics[width=1\textwidth]{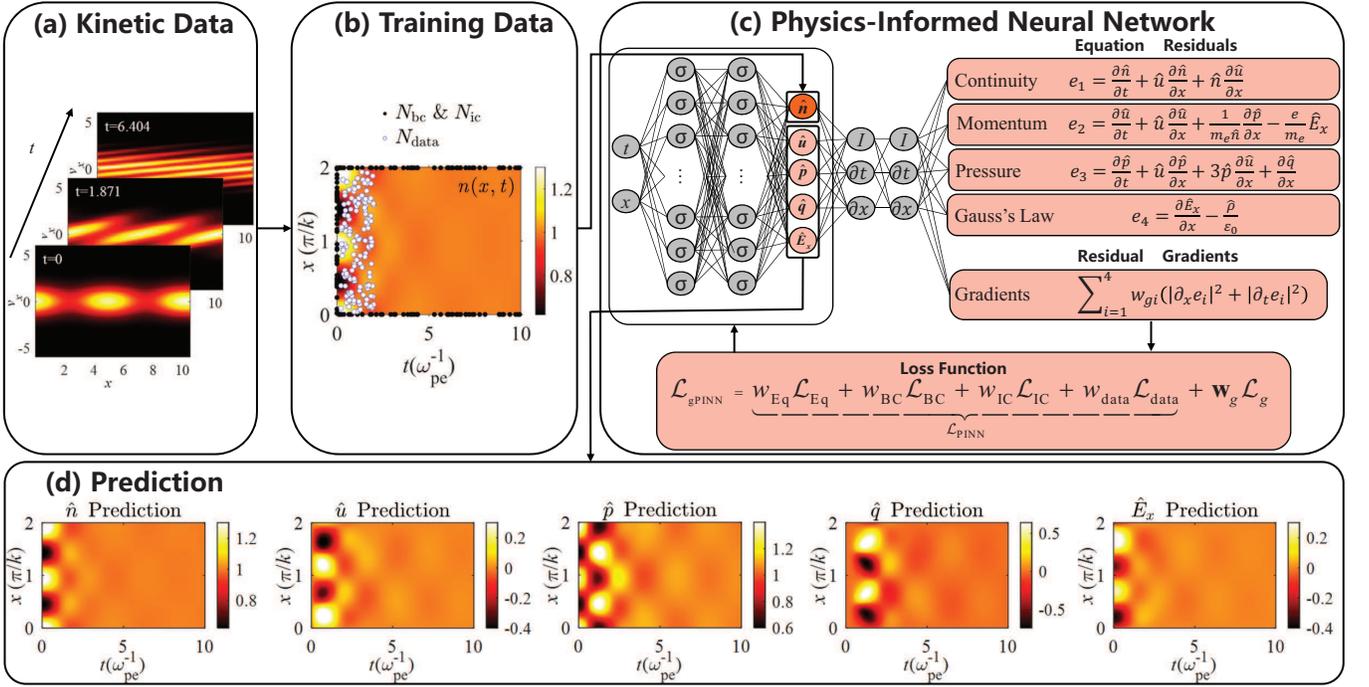}
\caption{Physics-informed neural network (PINN) architecture for the multi-moment fluid model with an implicit fluid closure learned from the kinetic simulation data. The whole procedure includes (a) kinetic simulation data generation, (b) sparse sampling of training data, (c) PINN
construction with the constraints of different moment equation residuals and their gradients, and (d) parameter prediction.}
\label{PINN construct}                
\end{figure*}

\section{Methodology}
\label{Methodology}
\subsection{Physical model}
\label{physics model}
Consider a collisionless plasma in the absence of a magnetic field, the dynamics of the plasmas are governed by the Vlasov equation which describes the evolution of the particle distribution function in the phase space $(\mathbf{r}, \mathbf{v})$, 
\begin{equation}
\frac{\partial f_{\mathrm{s}}}{\partial t}+\mathbf{v_s} \cdot \nabla_{\mathbf{r}} f_{s}+\left(\frac{e_{s}}{m_{s}} \right)\mathbf{E} \cdot \nabla_{\mathbf{v}} f_{s}=0
\end{equation}
where $f_{s}(\mathbf{r}, \mathbf{v_s}, t)$ is the velocity distribution function of particle species $s$ in a plasma, $e_{s}/m_{s}$ is the charge-to-mass ratio of the particle species $s$, and the operators $\nabla_{\mathbf{r}}=\left(\partial_{x}, \partial_{y}, \partial_{z}\right)$ and  $\nabla_{\mathbf{v}}=\left(\partial_{v_x}, \partial_{v_y}, \partial_{v_z}\right)$ are the gradient operators in configuration space and velocity space, respectively. For simplicity, we consider a one-dimensional model in $x$-$v_x$ space.
Additionally,  $E_{x}(x, t)$ is the self-induced electric field, which satisfies the Poisson equation describing the electrostatic field:
\begin{equation}
E_{x}(x, t)=-\nabla \phi
\label{poisson1}
\end{equation}
\begin{equation}
\bigtriangleup \phi= -\frac{\rho}{\varepsilon_{0}}
\end{equation}
Here, ${\phi}(x, t)$ is the electric potential, $\varepsilon_{0}$ is the vacuum permittivity, and ${\rho}(x, t)$ denotes the charge density:
\begin{equation}
\rho=\sum_{s} e_{s} n_{s}
\end{equation}
where $e_{s}$ and $n_{s}$ are the charge and number density of the particle species $s$, respectively. 

In general, Vlasov models tend to become more memory-consuming and computationally demanding due to the high dimensionality of phase space; so we consider fluid models of plasma that involve only the evolution of macroscopic quantities. Consequently, we obtain some macroscopic fluid quantities by calculating the moments of ${f}_{s}(x, v_s, t)$ in the velocity space, and then extract the evolution of the moments from the Vlasov simulation data. In detail, macroscopic fluid quantities including the number density $n_{s}(x,t)$, the fluid velocity $u_{s}(x,t)$, the pressure $p_{s}(x,t)$, and the heat flux ${q}_{s}(x,t)$ can be derived from the first three moment equations:
\begin{equation}
n_{s}(x,t) = \int {f_{s}(x,v_s,t)dv_s}
\label{eq_n}
\end{equation}
\begin{equation}
u_{s}(x,t) = \frac{1}{{n_{s}(x,t)}}\int {v_s f_{s}(x,v_s,t)dv_s}
\label{eq_u}
\end{equation}
\begin{equation}
{p}_{s}(x,t) = m_{s} \int {{{(v_s - u_{s})}^2}f_{s}(x,v_s,t)dv_s}
\label{eq_p}
\end{equation}
\begin{equation}
{q}_{s}(x,t) = m_{s} \int {{{(v_s - u_{s})}^3}f_{s}(x,v_s,t)dv_s}
\label{eq_q}
\end{equation}
The set of multi-moment fluid PDEs for electron species, $e$, is expressed as follows (we drop the subscript, $s$, for the variables hereafter for brevity):  
\begin{equation}
\frac{{\partial n_{}}}{{\partial t}} + u_{}\frac{{\partial n_{}}}{{\partial x}} + n_{}\frac{{\partial u_{}}}{{\partial x}} = 0
\label{ns}
\end{equation}
\begin{equation}
\frac{{\partial u_{}}}{{\partial t}} + u_{}\frac{{\partial u_{}}}{{\partial x}} + \frac{1}{{m_{e}n_{}}}\frac{{\partial p_{}}}{{\partial x}} = \frac{e}{m_{e}}{E_x}
\label{us}
\end{equation}
\begin{equation}
\frac{{\partial p_{}}}{{\partial t}} + u_{}\frac{{\partial p_{}}}{{\partial x}} + 3p_{}\frac{{\partial u_{}}}{{\partial x_{}}} + \frac{{\partial {q}_{}}}{{\partial x_{}}} = 0
\label{ps}
\end{equation}
\begin{equation}
\frac{\partial E_{x}}{\partial x}=\frac{\rho}{\varepsilon_{0}}
\label{Ex}
\end{equation}
These are the electron continuity, momentum, pressure, and Gauss's law equations, respectively. Clearly, the update of the lower-order moment equations (e.g., the pressure equation) depends on the evolution of the next-higher-order moment (e.g., the heat flux $q$); therefore, a comprehensive multi-moment fluid model must include a closure relation to close the system of equations. Because of the absence of the evolution of the fourth-order moment equation (or the heat flux equation) in the preceding equations, a closure relation for the heat flux $q$ is required for the multi-moment fluid model. 

\subsection{PINN and gradient-enhanced PINN architectures}\label{PINN theory}
The schematic diagram of the whole architecture depicted in Fig.~\ref{PINN construct} includes four parts: kinetic simulation data generation [Fig.~\ref{PINN construct}(a)], sparse sampling of the training data [Fig.~\ref{PINN construct}(b)], physics-informed neural network construction [Fig.~\ref{PINN construct}(c)], and parameter prediction [Fig.~\ref{PINN construct}(d)]. Beginning with the generation of kinetic simulation data by numerically solving the equations of the Vlasov-Poisson system, as depicted in Fig.~\ref{PINN construct}(a), we take snapshots of the velocity distribution $f(x,v_x)$ in phase space at several time steps to characterize these data. Secondly, for boundary and initial conditions, all physical variables $(n, u, p, q, E_x)$ are sampled, but only density $n$ is also sparsely sampled from the simulation data at the first few time steps as hypothetical known observations to train the neural network as shown in Fig.~\ref{PINN construct}(b) (or Fig.~\ref{sampling_diag}). Finally, the neural network with multi-moment fluid equation residual constraints is built to recover and forecast the number density $\hat n(x,t)$, the fluid velocity $\hat u(x,t)$, the pressure $\hat p(x,t)$, the heat flux $\hat{q}(x,t)$, and the electric field $\hat E_{x}(x,t)$ across the entire spatial and temporal range [Figs.~\ref{PINN construct}(c) and ~\ref{PINN construct}(d)]. A schematic illustration of the proposed PINN and gPINN, composed of a fully connected feedforward neural network (FNN) with multiple hidden layers and a residual network with the fluid moment equation and their gradient constraints, is depicted in Fig.~\ref{PINN construct}(c).

For the one-dimensional fluid model described by Eqs.~(\ref{ns})$-$(\ref{Ex}) on a spatial domain $\Omega\subset \mathbb{R}$, we first define the multi-moment fluid system deduced from the kinetic simulation data as the set ${\boldsymbol{\mathcal{F}}}\left ( x,t\right )=\{{n}_{}(x,t), {u}_{}(x,t), {p}_{}(x,t), {q}_{}(x,t), {E}_{x}(x,t)\}$ and then construct a neural network with the trainable parameters $\boldsymbol\theta$ to approximate the solution. The neural network as a parametric function approximator can be represented by a nonlinear function:
\begin{equation}
\begin{aligned}
\{\hat{n_{}}(x,t), \hat{u_{}}(x,t), \hat{p_{}}(x,t), \hat{{q}_{}}(x,t), \hat{E_{x}}(x,t)\}={\hat{\boldsymbol{\mathcal{F}}}}\left ( x,t;\boldsymbol\theta\right ),\\x \in\Omega, t \in[0, T]
\end{aligned}
\end{equation}
where $\boldsymbol\theta=\left\{\boldsymbol{W}, \boldsymbol{b}\right\}$ is the weight matrix and the bias vector. We take the derivatives of $\hat{\boldsymbol{\mathcal{F}}}$ with respect to $x$ and $t$ by applying automatic differentiation. PINN encodes professional physical priors into the loss function. These physical priors, which are expressed as a set of PDEs with appropriate initial and boundary conditions, are highly condensed knowledge of physical mechanisms that can inform the neural network. Then we utilize the constraints implied by the PDEs, the initial conditions, the boundary conditions, and some extra measurements of the density $n$ as labeled data to train the neural network. The whole loss function is defined as follows:
\begin{equation}
\begin{split}
{{\mathcal L}_{PINN}}&={w_{Eq}} \mathcal{L}_{Eq}+w_{BC} \mathcal{L}_{BC}\\
&+w_{IC} \mathcal{L}_{IC}+w_{data} \mathcal{L}_{data} 
\end{split}
\label{PINN_loss}
\end{equation}
where $w_{Eq}$, $w_{BC}$, $w_{IC}$ and $w_{data}$ are the weights of each loss function respectively. In this paper, we choose the weights $w_{Eq}=w_{BC}=w_{IC}=w_{data}=1$. In particular, we seek to minimize the residuals of the fluid moment equations, which are given as follows:
\begin{equation}
\begin{aligned}
{e_1} &=\frac{{\partial \hat {{n_{}}}}}{{\partial t}} + \hat {{u_{}}}\frac{{\partial \hat {{n_{}}}}}{{\partial x}} + \hat {{n_{}}}\frac{{\partial \hat {{u_{}}}}}{{\partial x}}\\
{e_2} &=\frac{{\partial \hat {{u_{}}}}}{{\partial t}} + \hat {{u_{}}}\frac{{\partial \hat {{u_{}}}}}{{\partial x}} + \frac{1}{{{m_{e}}\hat {{n_{}}}}}\frac{{\partial \hat {{p_{}}}}}{{\partial x}} - \frac{{e}}{{m_e}}\hat {{E_x}}\\
{e_3} &=\frac{{\partial \hat {{p_{}}}}}{{\partial t}} + \hat {{u_{}}}\frac{{\partial \hat {{p_{}}}}}{{\partial x}} + 3\hat {{p_{}}}\frac{{\partial \hat {{u_{}}}}}{{\partial {x_{}}}} + \frac{{\partial \hat {{{q}}}}}{{\partial {x}}}\\
{e_4} &=\frac{{\partial \hat {{E_x}}}}{{\partial x}} - \frac{{\hat \rho }}{{{\varepsilon _0}}}
\label{e1-e4}
\end{aligned}
\end{equation}
\begin{equation}
{\mathcal{L}_{Eq}} =  \frac{1}{{{N_{eq}}}}\sum\limits_{j = 1}^{{N_{eq}}} {\sum\limits_{i = 1}^4 {{{\left| {{e_i}\left( {x_j,t_j} \right)} \right|}^2}} }, x_{j}\in\Omega,t_{j}\in[0, T]
\label{Eq_loss}
\end{equation}
Here $e_1$ denotes the continuity equation residual, $e_2$ denotes the momentum equation residual, $e_3$ denotes the pressure equation residual, and $e_4$ denotes the Gauss's law equation residual. $N_{eq}$ is the number of trained data for $\mathcal{L}_{Eq}$. In fact, we want to conduct an inverse problem using PINN, where the fluid closure is implicitly included in the neural network, assuming that both the initial and boundary conditions are known and sparsely sampled,
\begin{equation}
\begin{aligned}
\mathcal{L}_{BC}=\frac{1}{N_{bc}} \sum_{j=1}^{N_{bc}}\sum_{i=1}^{5}\left|{\hat{\boldsymbol{\mathcal{F}}}}_{i}\left(x_{j}, t_{j};\boldsymbol\theta\right)-{\boldsymbol{\mathcal{F}}}_{i}\left(x_{j}, t_{j}\right)\right|^{2},\\
x_{j}\in\partial \Omega,t_{j}\in[0, T]
\end{aligned}
\end{equation}
\begin{equation}
\begin{aligned}
\mathcal{L}_{IC}=\frac{1}{N_{ic}} \sum_{j=1}^{N_{ic}}\sum_{i=1}^{5}\left|{\hat{\boldsymbol{\mathcal{F}}}}_{i}\left(x_{j}, t_{j};\boldsymbol\theta\right)-{\boldsymbol{\mathcal{F}}}_{i}\left(x_{j}, t_{j}\right)\right|^{2},\\
x_{j}\in\Omega,t_{j}=0
\end{aligned}
\label{loss_IC}
\end{equation}

Although the model inputs should ensure that there is enough information for the neural network to accurately capture the governing equations of the system, the amount of input information should be minimized. Therefore, we only sample the kinetic simulation data in the first few time steps as labels (see Fig.~\ref{sampling_diag}) to allow the network to capture the fluid closure that incorporates the kinetic effects. In this paper, only the density $n$ is sampled.
\begin{equation}
\begin{aligned}
\mathcal{L}_{data}=\frac{1}{N_{data}} \sum_{j=1}^{N_{data}}\left|\hat{n}\left(x_{j}, t_{j}\right)-{n}\left(x_{j}, t_{j}\right)\right|^{2},\\
x_{j}\in\Omega,t_{j}\in[0, t^{\prime}],t^{\prime} \le \frac{T}{5} 
\end{aligned}
\end{equation}

Meanwhile, other studies have demonstrated that gPINN improves the accuracy of PINN, especially when applied to PDEs with steep gradients~\cite{yu2022gradient}. Thus, we introduce gPINN to capture the structures with large gradients. The main idea of gPINN embeds the gradient information into the loss function by enforcing that the derivatives of the moment equation residuals be the minimum. Assuming that the gradient of the equation residual $\nabla \boldsymbol{e}$ exists, the loss function of gPINN is:
\begin{equation} 
\begin{split}
{{\mathcal L}_{gPINN}}&={w_{Eq}} \mathcal{L}_{Eq}+w_{BC} \mathcal{L}_{BC}\\
&+w_{IC} \mathcal{L}_{IC}+w_{data} \mathcal{L}_{data} + {{\bf{w}}_g}{\mathcal{L}_g}
\end{split}
\label{gPINN_loss}
\end{equation}
For the $1X1V$ case,  the additional loss term is,
\begin{equation}
\begin{aligned}
\mathbf{w}_g{{\cal L}_g} = \frac{1}{{{N_g}}}\sum\limits_{j = 1}^{{N_g}} {\sum\limits_{i = 1}^4 {w_{gi}\left( {{{\left| {{\partial _x}{e_i}\left( {x_j,t_j} \right)} \right|}^2} + {{\left| {{\partial _t}{e_i}\left( {x_j,t_j} \right)} \right|}^2}} \right)} },\\
x_{j}\in\Omega,t_{j}\in[0, T]
\label{gPINNrd}
\end{aligned}
\end{equation}
The weight $\mathbf{w}_g=\left \{ w_{g_{1}},w_{g_{2}},w_{g_{3}},w_{g_{4}}\right \}$ is an extra hyperparameter in the gPINN architecture for optimization. 

Conventional gPINN architectures incorporate the gradient terms of all the equation residuals and add them to the loss function~\cite{yu2022gradient}. In this paper, we make an attempt to only include the gradient of a specific equation residual as the additional constraint, which also reduces the computational cost compared with the traditional gPINN. Here, we define a variant of gPINN that only includes the gradient of the pressure equation residual, namely, gPINN$p$. The idea of gPINN$p$ is motivated by the fact that the heat flux $q$ and the pressure $p$ are closely related in the residual $e_3$ of Eq.~(\ref{e1-e4}). 

\section{Simulation} 
\label{Simulation}

\subsection{Synthetic model setup}
\subsubsection{Kinetic simulation data generation}

This section describes the kinetic Vlasov-Poisson simulations used to generate the training data. The physical problem under investigation is Landau damping in a collisionless, electrostatic plasma. The initial setup consists of an immobile, neutralizing ion background and two perturbation modes applied to the electron density,
\begin{equation}
{n_e}(x,t = 0) = {n_0}(1 + A_{1} \cos \left(k_{1}x\right)+A_{2} \cos \left(k_{2} x+\varphi\right))
\label{eq_ne_inital}
\end{equation}
\begin{equation}
{n_i}(x,t = 0) = {n_0}
\label{eq_ni_inital}
\end{equation}
where $n_0$ is the initial density of each species, $k_{1}$ and $k_{2}$ are the wavenumbers of the two modes, $A_{1}$ and $A_{2}$ are their amplitudes, and $\varphi$ is a random phase. 

We use the open-source continuum Vlasov code \texttt{Gkeyll} \footnote{https://github.com/ammarhakim/gkyl} for this study. The simulation employs a periodic configuration domain, $0 < x <\frac{2\pi}{k_{1}}$, discretized to 128 cells, and a plasma velocity space ${\rm{ - }}6{v_{th_s}} < v_x < 6{v_{th_s}}$ with 128 cells. A fixed time step size $\Delta t=0.001\omega_{pe}^{-1}$ is used and the simulation takes 10000 steps before it stops at $t=10\omega_{pe}^{-1}$.
The numerical scheme being used is a discontinuous Galerkin method with second-order serendipity polynomial bases~\cite{Juno2018}.
The specific simulation parameters are summarized in Table~\ref{tab1}.  

To construct the training datasets, the electron density $n$, velocity $u$, pressure $p$, and heat flow $q$ are extracted from the phase-space data following Eqs.~(\ref{eq_n})-(\ref{eq_q}).

\begin{table}[h!t]
\scriptsize
\centering
\caption{Summary of the initial setup parameters}
\label{tab1}

\begin{tabular}{ccccc}\midrule\midrule
$k_{1}\qquad\quad$ & $k_{2}\qquad\quad$ & $A_{1}\qquad\quad$ & $A_{2}\qquad\quad$ & $\varphi$ \\\midrule
0.6 $\qquad\quad$ & 1.2 $\qquad\quad$ & 0.05 $\qquad\quad$ & 0.4 $\qquad\quad$ & 0.38716 \\\midrule\midrule
\end{tabular}

\end{table}

Here we want to point out that the damping rates of these two modes are different; the mode with a short wavelength predominates but decays fast, and the one with a long wavelength of low energy decays slowly.

\subsubsection{Deep neural network setup}

Both PINN and gPINN involve neural network architecture selection since it has a significant impact on the prediction precision. The parameters of the neural network utilized are shown in Table~\ref{tab2}. All of them were determined by trial and error while taking into account the solution precision, convergence, and computational efficiency.
\begin{table}[h!t]
\scriptsize
   \centering
   \caption{Parameter setting of PINN and gPINN (and gPINN$p$)}
   \label{tab2}
  \resizebox{\linewidth}{!}{ 
    	\begin{tabular}{c c c c c}\midrule\midrule
        \makecell[c]{Numbers of layers and \\  neurons within hidden layer} & Optimizer & \makecell[c]{Learning \\ rate} & \makecell[c]{Activation\\ function} & {Batch size} \\\midrule
        5 and 50  & ADAM & \makecell[c]{0.01} & Swish & 10000\\
    \midrule\midrule
	\end{tabular}
	}
\end{table}

The artificial neural network consists of the input layer, the hidden layer, and the output layer. A hidden layer that contains five layers and 50 neurons provides the most accurate solutions. In addition, we choose this network structure in order to reduce additional computational requirements or overfitting. We choose a non-linear activation function Swish~\cite{Ramachandran2017SwishAS} to retain information about the gradient of the data with respect to the input variables ($x$ and $t$). In each iteration of the ADAM optimizer~\cite{kingma2014adam}, the mini-batch of data and the residual points used to penalize the equation are processed and have a size of 10000. To avoid an unstable training process caused by a rapid change in the learning rate, we use a constant learning rate of 0.01. In addition, we utilize weight normalization to accelerate the training of PINN (and gPINN)~\cite{salimans_weight_nodate}.

\begin{figure}[htbp]
 \includegraphics[width=0.5\textwidth]{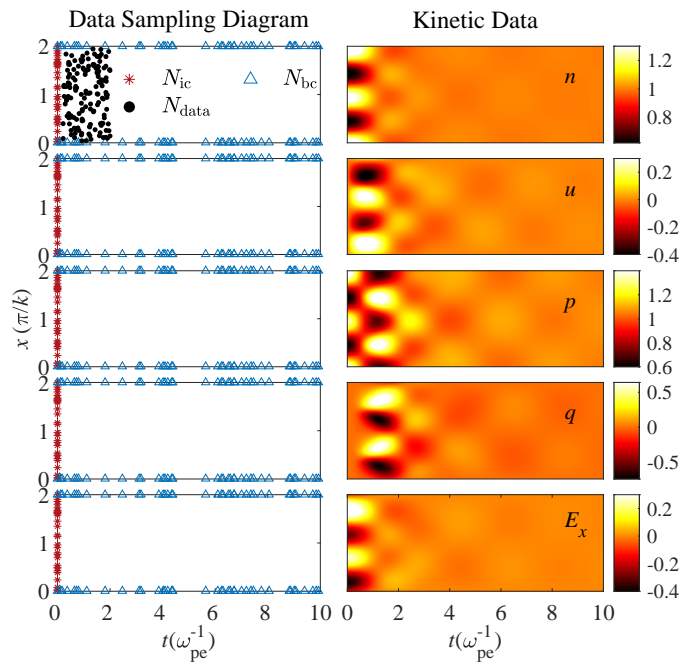} 
    \caption{Left: Schematic diagrams of training data sampling from
the simulation domain. Right: From top to bottom, the kinetic simulation
data of density $n$, velocity $u$, pressure $p$, heat flux $q$, and
electric field $E_x$ , respectively.}                    
\label{sampling_diag}  
\end{figure}

Data sampling diagrams of the neural network are displayed in Fig.~\ref{sampling_diag}. The training data are composed of the randomly sampled  $N_{ic} = 200$ initial conditions and $N_{bc} = 300$ boundary points from the five quantities, i.e., $n$, $u$, $p$, $q$, and $E_x$. For the electron density $n$, we also sample a small set of points from 0 to 2$\omega_{pe}^{-1}$ as extra measurements, totaling $N_{data}=23863$ sampling points, which correspond to a sampling rate of approximately $3.125\%$ within $t^{\prime}=2\omega _{pe}^{-1}$. 

\begin{figure*}[htbp]
\includegraphics[width=1.0\textwidth]{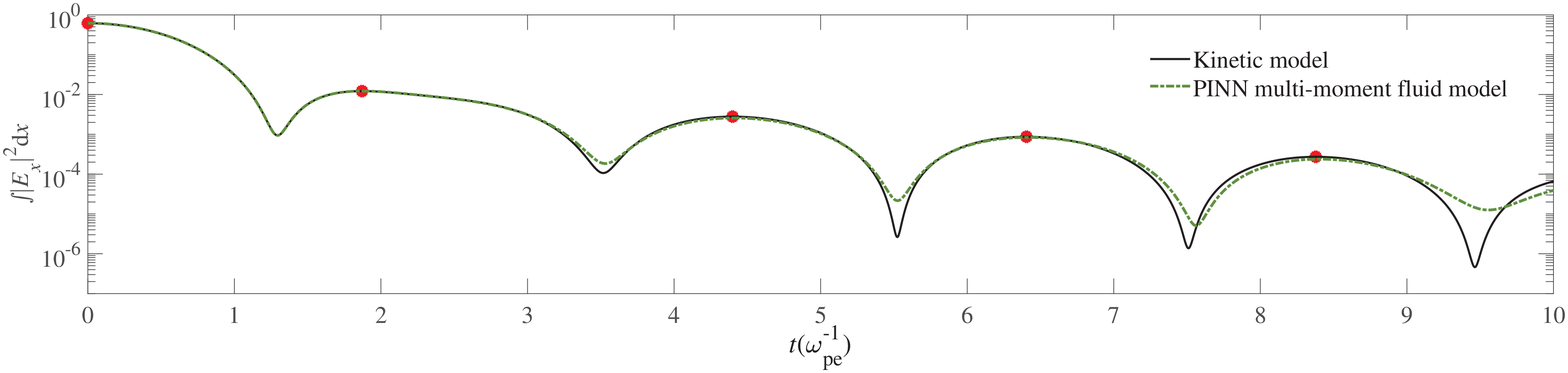}
\includegraphics[width=1.0\textwidth]{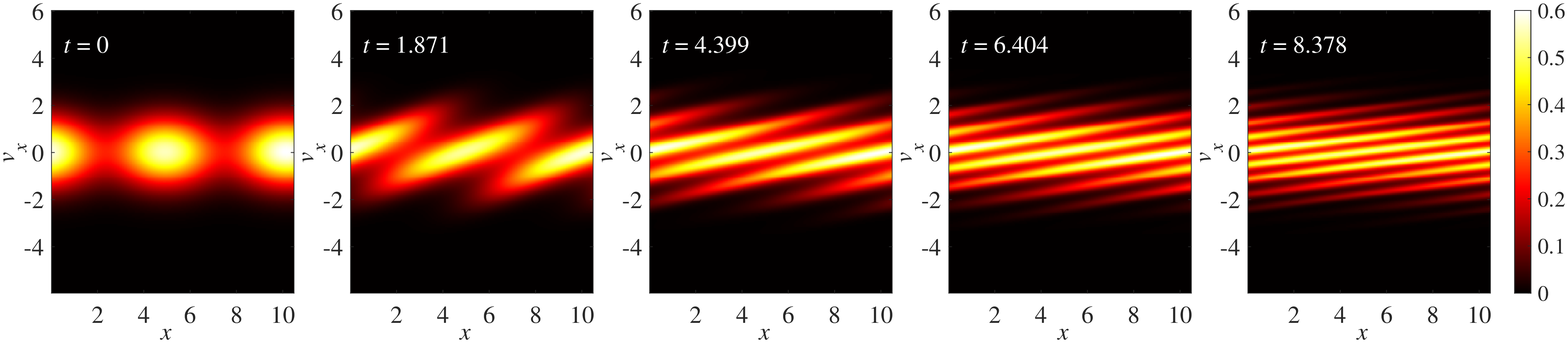}
\caption{Top: The evolution of the electric field energy predicted by the multi-moment fluid model constructed using PINN (green dashed line) and the kinetic model (black solid line), with wave peaks at times $t= 0$, $1.871\omega_{pe}^{-1}$, $4.399\omega_{pe}^{-1}$, $6.404\omega_{pe}^{-1}$, and $8.378\omega _{pe}^{-1}$ (red dots). Bottom: The velocity distribution $f(x,v_x)$ in phase space at these peak points (see red dots in the top panel).}
\label{velocity distribution}
\end{figure*}

\subsection{Data-driven modeling results} 
Based on the kinetic simulation data, we construct the multi-moment fluid model using PINN and gPINN, respectively. When training is converged, the neural network simultaneously predicts the values of $\hat{n}$, $\hat{u}$, $\hat{p}$, $\hat{{q}}$, and $\hat{E_x}$ for the whole time period up to $10 {\omega _{pe}^{-1}}$. We use the integral of the electric field square over the entire configuration space $\int\left|E_{x}\right|^{2} d x$ to evaluate the accuracy of PINN (and gPINN) on capturing the Landau damping process.

Figure~\ref{velocity distribution} shows the evolution of the electric field energy predicted by the PINN-constructed multi-moment fluid model over time, and the temporal evolution of velocity phase space distribution $f(x,v_{\rm x})$ at several fixed time steps from the kinetic simulation data. Based on the sparsely sampled electron density $n$ in the time period $0 < t < 2\omega _{pe}^{-1}$, the PINN-constructed multi-moment fluid model recovers and reconstructs the electric field energy evolution during the time period $t=[0, 10\omega_{pe}^{-1}]$. The predicted electric field energy oscillates and decays as time progresses, with wave peaks at times $t= 0, 1.871\omega _{pe}^{-1}, 4.399\omega _{pe}^{-1}, 6.404\omega _{pe}^{-1}$, and $8.378\omega_{pe}^{-1}$ (labeled as red dots), which agrees with the kinetic simulation data. The evolution of the complicated velocity distribution $f(x,v_{\rm x})$ in phase space from the kinetic simulation data at these time steps (labeled as red dots) is depicted in the bottom panels of Fig.~\ref{velocity distribution}. The good agreement between the kinetic simulation data and PINN-generated data indicates that the PINN-constructed multi-moment fluid model is capable of accurately representing the complicated evolution of the plasma dynamics and capturing the Landau damping process even without directly evolving the distribution function in the phase space.

For quantitative assessment, we define the absolute error (AE) as the evaluation metrics, which is expressed as:
\begin{equation}
    \operatorname{AE}(\hat{y}, {y})=|\hat{y}(x,t)-{y}(x,t)|
\end{equation}
where $\operatorname{AE}(\hat{y}, {y})$ is defined as the difference between the outputs ${\hat{y}(x,t)}$ of the neural networks and the kinetic simulation data ${y}(x,t)$.

\begin{figure}[htbp]
\centering
\subfloat{a$) $\,PINN}{
{\includegraphics[width=0.49\textwidth]{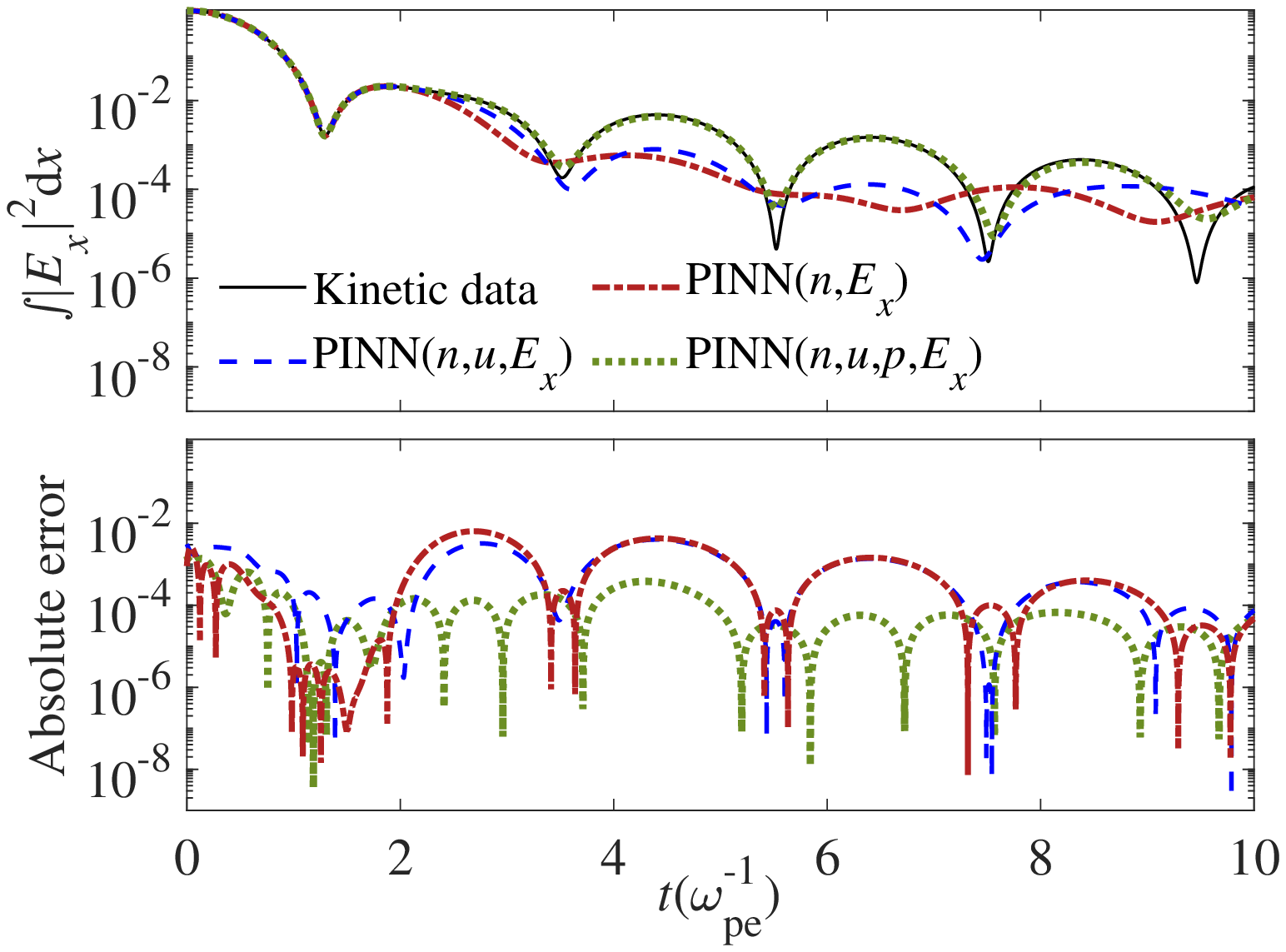}}}
\centering

\subfloat{b$) $\,gPINN}{
{\includegraphics[width=0.49\textwidth]{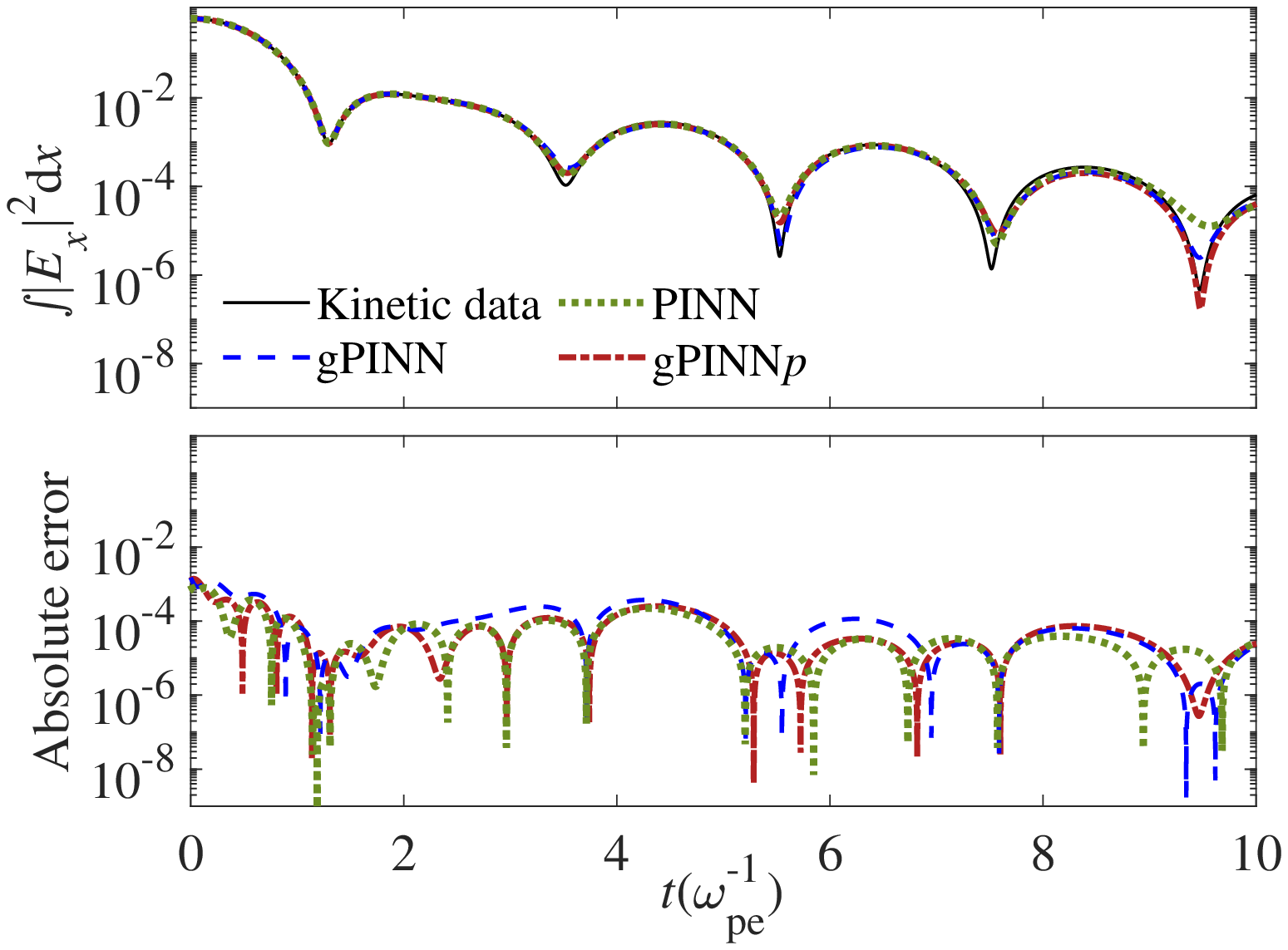}}}
\caption{Temporal evolution of electric field energy and absolute error using (a) PINN and (b) gPINN with different moment equation residuals as constraints. The weights $w_{g i}=0.01~(i = 1,2,3,4)$ and $w_{g 3}=0.01$ are adopted for gPINN and gPINN$p$, respectively, following hyperparameter tuning.}
\label{PINN_diff_epu}
\end{figure}

The introduction of equation residuals into the loss function [see Eq.~(\ref{Eq_loss})] is the most important component of our scheme, while the equations provide the necessary physical priors and serve as a roadmap for network optimization. It is noteworthy that our method does not involve any explicit fluid closure equations, but the fluid closure relation is implicitly included in the neural network. 

In Fig.~\ref{PINN_diff_epu}(a), we compare the performances of the PINN-constructed multi-moment fluid model using different combinations of moment equation residual as constraints [see Fig.~\ref{PINN construct}(c) or Eq.~(\ref{e1-e4})]. The goal is to determine the minimum requirement to accurately capture the Landau damping process. In more detail, the electric field equation residual is always retained, and the number of equation residuals is always 1 less than the number of network predictions. As an example, PINN($n$, $u$, $E_x$) in Fig.~\ref{PINN_diff_epu}(a) indicates the use of continuity and momentum equation residuals, as well as the electric field equation residual, while the model outputs are $\hat{n}$, $\hat{u}$, $\hat{p}$, and $\hat{E_{x}}$. In order to compare and show more clearly the differences between the results, we give the absolute errors between the predictions using different numbers of equation residuals and the kinetic simulation data. The overall absolute error of PINN($n$, $u$, $p$, $E_x$) is less than $10^{-3}$, which is smaller than the other two cases using fewer equation residuals, i.e., PINN($n$, $E_x$) and PINN($n$, $u$, $E_x$). 

The results obtained by the PINN-constructed multi-moment fluid model in Fig.~\ref{PINN_diff_epu}(a) without using the pressure equation residual as a constraint have seriously deviated from the true value (or the kinetic simulation data). Consequently, we draw the conclusion that it is necessary to use at least the first three moment equations as constraints and such a fluid system contains  five variables ($n$, $u$, $p$, $q$, and $E_x$) to accurately capture the Landau damping process. When this condition is not satisfied, i.e., the number of constraints is less than the minimum requirement, the PINN-constructed multi-moment fluid model is not able to capture the specific kinetic effects due to the lack of sufficient input information. Meanwhile, Fig.~\ref{PINN_diff_epu}(a) also demonstrates that the results begin to deteriorate at later stages, particularly in the wave troughs of the electric field energy curve, where large deviations are observed. Therefore we adopt gPINN by adding the gradients of the moment equation residuals, which has been demonstrated to be more effective than PINN~\cite{yu2022gradient} when addressing similar issues.

The performances of the PINN-constructed and gPINN-constructed multi-moment fluid models in capturing the Landau damping process are compared in Fig.~\ref{PINN_diff_epu}(b). Here, we introduce a variant of gPINN, namely, gPINN$p$, which only incorporates the gradient of the pressure equation residual as an additional constraint, while the traditional gPINN includes the gradients of all the equation residuals [see Fig.~\ref{PINN construct}(c) or Eq.~(\ref{gPINNrd})]. The idea of gPINN$p$ is motivated by the fact that the heat flux $q$ and the pressure $p$ are closely related in the residual $e_3$ of Eq.~(\ref{e1-e4}). Meanwhile, gPINN$p$ is computationally cheaper than the traditional gPINN with more constraints. In both gPINN and gPINN$p$, the gradient weight $w_{gi}$ is a hyperparameter, filtered by the optimized tests with $w_{gi} = 0.01$ ($i = 1,2,3,4$ for gPINN and $i =3$ for gPINN$p$). In Fig.~\ref{PINN_diff_epu}(b), the absolute error of the electric field energy between predicted and true values reaches $2.22\times10^{-5}$ using PINN, $2.00\times10^{-6}$ using gPINN, and $4.68\times10^{-7}$ using gPINN$p$ at the last wave trough approximately at time $t=9.5\omega _{pe}^{-1}$. The multi-moment fluid model constructed using standard gPINN fits the kinetic simulation data better than that using PINN, while the gPINN$p$-constructed multi-moment fluid model provides the most refined results, especially at later stages $t>8\omega _{pe}^{-1}$ when the electric field energy decays to relatively low values. The finding that the gPINN$p$-constructed multi-moment fluid model has the best performance indicates that the evolution of the heat flux $q$ heavily relies on the pressure $p$ and its gradients, consistent with the theoretical expectation~\cite{Hunana2019}. 

\begin{figure}[htbp]
\centering
\includegraphics[width=0.49\textwidth]{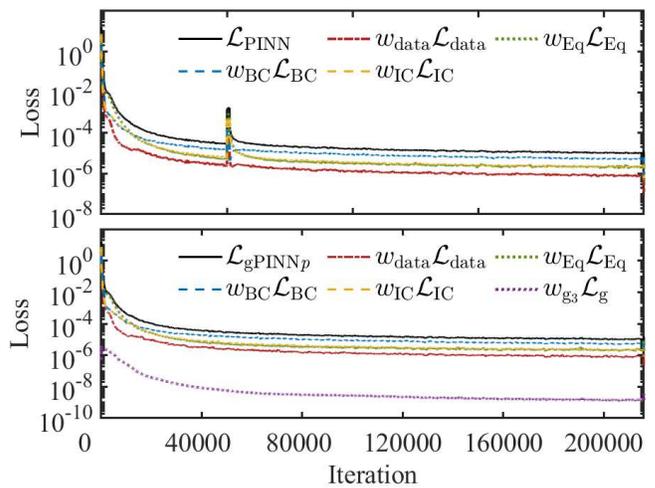}
\caption{History of the aggregate losses and loss of various components of PINN (top) and gPINN$p$ with $w_{g_3}=0.01$ (bottom). A moving average filter with a moving window of length 100 time iterations is used to smooth losses.}
\label{Loss}
\end{figure}

In Fig.~\ref{Loss}, we record the aggregate losses ${{\mathcal L}_{PINN}}$ and ${{\mathcal L}_{gPINN}}$ of PINN and gPINN$p$ during the whole training process, as well as each component of the loss function. To filter out the oscillations in the time series for the loss values, we employ a centered moving average by sliding a window of length 100 iterations. ${{\mathcal L}_{PINN}}$ and ${{\mathcal L}_{gPINN}}$ showed a general downward trend, with a gentle trend after 40,000 iterations and they converge to roughly $10^{-5}$ after a total of 216000 iterations. For the definition of the loss function for PINN and gPINN$p$, see Eqs.~(\ref{PINN_loss}) and (\ref{gPINN_loss}), respectively. As for the training procedure, all convergent results are obtained after $2\times 10^{5}$ steps of iterative optimization. Among various contributing components to ${{\mathcal L}_{PINN}}$ and ${{\mathcal L}_{gPINN}}$, the smallest is the data loss, ${{\mathcal L}_{data}}$, followed by the equation residual loss and the loss at initial conditions, ${{\mathcal L}_{Eq}}$ and ${{\mathcal L}_{IC}}$, respectively. In contrast, the boundary condition loss, ${{\mathcal L}_{BC}}$, has the largest magnitude. For $\mathcal{L}_{gPINNp}$, the contribution of the gradient loss, $\mathcal{L}_{g}$, remains relatively small due to its small weight $\omega_g$, which is determined through hyperparametrization. This small contribution is critical, though, for achieving better performance.

\begin{figure*}[htbp]
    \centering
 \includegraphics[width=0.82\textwidth]{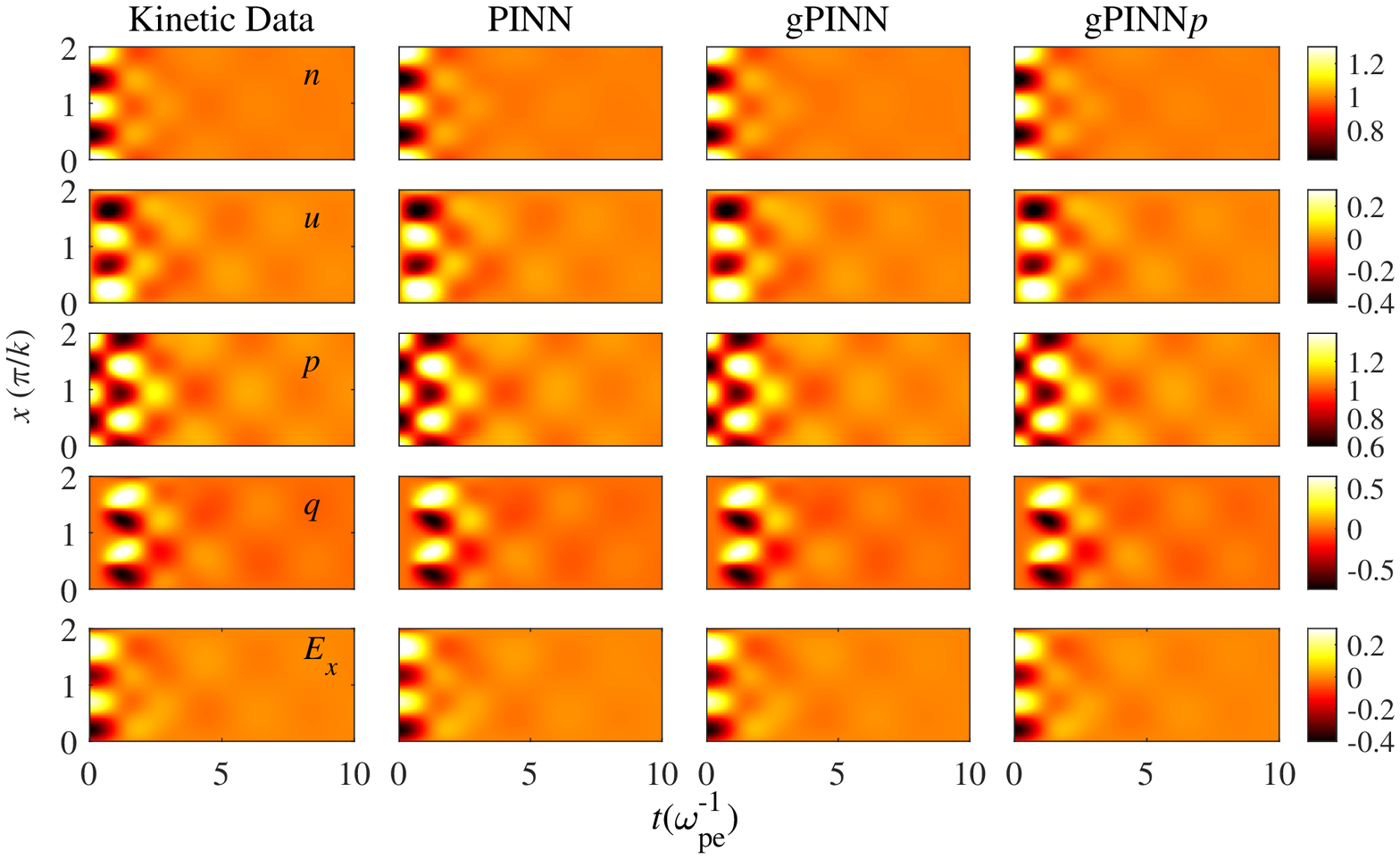}
    \caption{Comparison of the predicted physical quantities (from the multi-moment fluid models constructed using PINN, gPINN, and gPINN$p$) with respect to the kinetic simulation data. Each panel from top to bottom shows the density $n$, velocity $u$, pressure $p$, heat flux ${q}$, and electric field $E_x$, respectively.}
    \label{fig:compare-pinn-gpinn}
\end{figure*}

\begin{figure*}[htbp]
    \centering
 \includegraphics[width=0.82\textwidth]{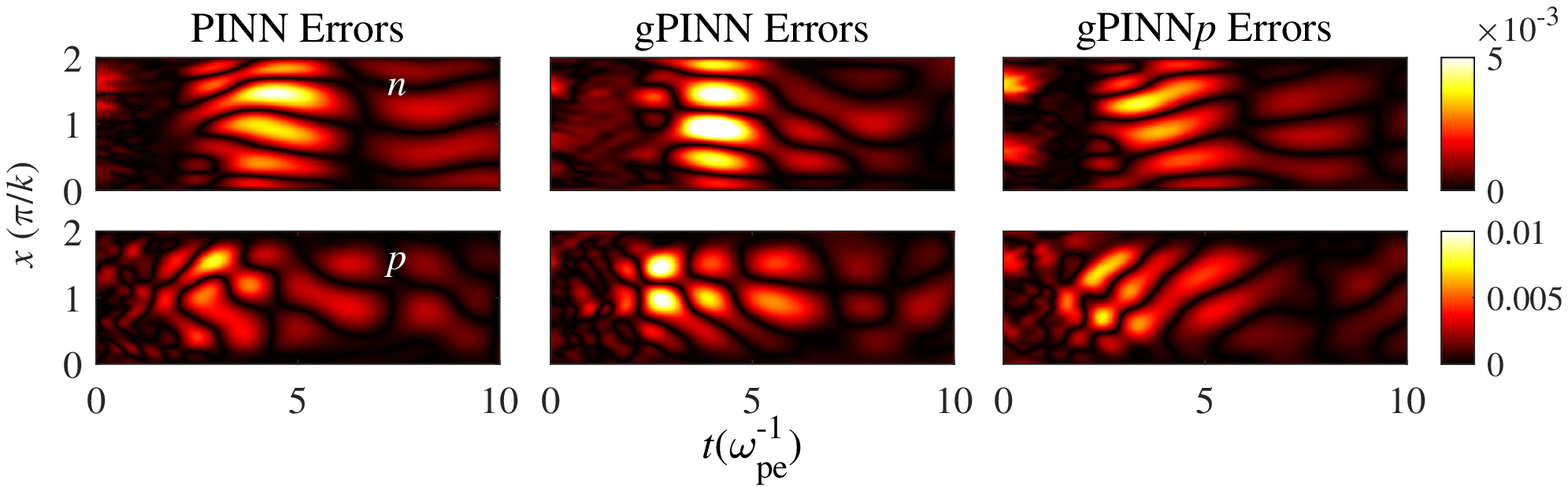} 
    \caption{Comparison of the relative errors between the predicted physical quantities (from the multi-moment fluid models constructed using PINN, gPINN, and gPINN$p$) and the kinetic simulation data. The top and bottom rows show the relative errors of the density $n$ and pressure $p$, respectively.}
    \label{fig:error-pinn-gpinn}
\end{figure*}

In Fig.~\ref{fig:compare-pinn-gpinn}, we present the temporal-spatial evolution of the physical quantities predicted by the multi-moment fluid models constructed using PINN, gPINN, and gPINN$p$, and the true value from the kinetic simulation. The corresponding row in each column displays the relevant values for the density $n$, velocity $u$, pressure $p$, heat flux ${q}$, and electric field $E_x$, respectively. As shown in Fig.~\ref{fig:compare-pinn-gpinn}, the PINN, gPINN, and gPINN$p$ architectures have the ability to accurately reconstruct and predict those physical quantities with an implicit fluid closure included in the neural network. It is noteworthy that the accurate prediction of these quantities only relies on sparse sampling of a small fraction of the kinetic simulation data (see Fig.~\ref{sampling_diag}). Most importantly, the neural network not only captures the kinetic damping of integral electrostatic energy but also reproduces the spatial-temporal profile of the physical quantities.
 
Figure~\ref{fig:error-pinn-gpinn} depicts the relative errors (RE) of the predicted quantities from the multi-moment fluid models constructed with PINN, gPINN, and gPINN$p$ based on the definition, $\operatorname{RE}(\hat{y}, {y})=\left|\frac{\hat{y}(x, t)-y(x, t)}{y(x, t)}\right|$, where ${y}(x,t)$ are the kinetic simulation data ${\hat{y}(x,t)}$ are the neural network outputs. All physical quantities predicted by the three models are in good agreement with kinetic simulation data. This particular comparison does not seem to clearly favor any neural network. Note that the relative errors due to each run were computed against the ``true'' solution frame by frame in time. Thus the errors may be contaminated by the subtle phase errors of the neural network predictions and do not necessarily reflect the true performance of the models. Such anomalous errors would be particularly distracting when computing relative errors in quantities fluctuating near zero values, such as $u$, $q$, and $E_x$, since the denominator may vanish, making direct comparison extremely difficult. Hence the relative errors of these terms are not shown in Fig.~\ref{fig:error-pinn-gpinn}. Nevertheless, based on previous analysis and considerations, the gPINN$p$ architecture exhibits superior performance compared with the other architectures.
 
\section{Conclusion and Discussion}
\label{Conclusion and Discussion}
In conclusion, we construct multi-moment fluid models using PINN and gPINN, where the fluid closure is learned from the kinetic simulation data and is implicitly included in the neural networks. The neural networks use the physical constraints of the multi-moment fluid equation residuals and their gradients. In order to accurately capture the Landau damping process, PINN and gPINN need to include the first three moment equations (i.e., equations of $n$, $u$, and $p$) as constraints. Meanwhile, the PINN and gPINN architectures are capable of accurately predicting all these physical quantities concurrently. 

In addition, we propose and explore a variant of gPINN, namely, gPINN$p$. Unlike the traditional gPINN that uses the gradients of all the moment equation residuals as additional constraints, the gPINN$p$ architecture only adopts the gradient of the pressure equation as the additional constraint. Compared with the results from the cases using PINN and gPINN, the gPINN$p$-constructed multi-moment fluid model provides the most accurate predictions, especially at later stages. The finding that gPINN$p$ has the best performance indicates that the evolution of the heat flux $q$ heavily relies on the pressure $p$ and its gradients, consistent with the theoretical expectation. 

In the future, we intend to expand the extrapolation capabilities of the neural networks in order to apply PINNs to higher-dimensional and more intricate multiscale problems.

\section*{Acknowledgments}
The authors from Fudan University were supported by the National Key Research and Development Program of China (2021YFA0717300) and National Science Foundation of China (42074189). C.D. was supported by the U.S. Department of Energy (through LDRD) under contract number DE-AC02-09CH11466.

\nocite{}
%

\end{document}